\documentstyle[aps,prb,epsf]{revtex}
\draft
\begin{document}
% Macros for the various macro package names, etc.
\def\SNG{{\em Physical Review Style and Notation Guide}}
\def\LUG {{\em \LaTeX{} User's Guide \& Reference Manual}}
\def\btt#1{{\tt$\backslash$\string#1}}%
\def\REVTeX{REV\TeX}
\def\AmS{{\protect\the\textfont2
        A\kern-.1667em\lower.5ex\hbox{M}\kern-.125emS}}
\def\AmSLaTeX{\AmS-\LaTeX}
\def\BibTeX{\rm B{\sc ib}\TeX}
%\makeatletter
%\tighten
\twocolumn[\hsize\textwidth\columnwidth\hsize\csname@twocolumnfalse%
\endcsname
\title{Magnetic pair breaking in disordered superconducting films}
\author{T. P. Devereaux}
\address{Department of Physics, University of California, Davis, CA 95616}
\author{D. Belitz}
\address{Department of Physics and Materials Science Institute, University of Oregon, Eugene, OR 97403}
\date{\today}
\maketitle
\begin{abstract}
A theory for the effects of nonmagnetic disorder on the magnetic 
pair breaking rate $\alpha$ induced by paramagnetic impurities 
in quasi two-dimensional superconductors is presented. Within the
framework of a strong-coupling theory for disordered superconductors,
we find that the disorder dependence of $\alpha$ is determined by the
disorder enhancements of both the electron-phonon coupling and the 
spin-flip scattering rate. These two effects have a tendency to
cancel each other. With parameter values appropriate for
Pb$_{0.9}$Bi$_{0.1}$, we find a
pair breaking rate that is very weakly dependent on
disorder for sheet resistances $0<R_{\Box} \alt 2.5\ {\rm k\Omega}$,
in agreement with a recent experiment by Chervenak and Valles.
\end{abstract}
%\receipt{ - draft - \today}
%\vspace{2.5 cm}
\pacs{PACS numbers: 74.20.Mn, 74.60.Ge, 64.60.Ak, 74.40.+k}
\vskip 1cm
]
%\narrowtext
\section{introduction}
\label{sec:I}

The physics of the observed $T_{c}$ suppression in superconductors
that contain nonmagnetic disorder has been the subject of much debate in
recent years. Let us focus on homogeneously disordered thin superconducting
films, with the disorder parametrized by the normal-state sheet resistance
$R_{\Box}$, and on the BCS-like quasi transition that is well pronounced
in these films although the true superconducting transition is of
Kosterlitz-Thouless nature.\cite{films} In these systems, the BCS transition
temperature $T_c$, defined as the mid-point of the resistive transition,
is observed to decrease monotonically with increasing disorder.\cite{films,R}
A complete quantitative understanding of this effect within a
microscopic strong-coupling theory has proven difficult, although
the first perturbative calculations within a phenomenological BCS
model\cite{fukuyama} were rather promising.
Qualitatively, disorder-induced changes in the electron-phonon 
coupling,\cite{schmid} in the Coulomb repulsion between the constituents
of the Cooper pairs,\cite{amr} and in the normal-state 
density of states\cite{dbdos} have all been identified to be important.
The difficulty lies in the fact that some of these effects tend to
suppress $T_c$ while others tend to enhance it, and $T_c$ depends
exponentially on all of them so that subtle balancing effects result.
Also, the number of parameters that acquire a disorder dependence is
quite large, and fits of theoretical results to experimental data are
therefore not necessarily very conclusive. Indeed, theories that are
structurally quite different, and mutually inconsistent, have been
shown to fit the same $T_c$ data equally well.\cite{f,trkdb}
 
In this situation it is obvious that one should study the disorder
dependence of quantities other than the transition temperature in an
attempt to discriminate between various theories, and to obtain
independent information about the disorder dependence of the various 
parameters that determine $T_c$. One possibility is to measure the
inelastic lifetime of quasiparticles.\cite{qp,pl} The experiment by
Pyun and Lemberger\cite{pl} on amorphous InO has been analyzed by the
present authors\cite{tpddb} in the framework of a strong-coupling theory
for disordered superconductors,\cite{db} and quantitative agreement
between theory and experiment has been achieved. Another possibility
is to study the influence of nonmagnetic disorder on the pairbreaking
induced by a small amount of magnetic impurities in addition to the
nonmagnetic ones. This has the advantage that the pair breaking parameter
is easier to measure than inelastic lifetimes, and that it can be
measured simultaneously with the $T_c$ suppression in a series of
films where the nonmagnetic disorder is varied {\it in situ} by
controlling the film thickness.

Such an experiment has recently been performed by Chervenak and 
Valles,\cite{jj} who studied quench condensed ultrathin films of 
Pb$_{0.9}$Bi$_{0.1}$ of varying degrees of 
disorder ($150\ \Omega < R_{\Box} < 2.2\ {\rm k}\Omega$, leading to a
$T_{c}$ between $6$\ K and $2.35$\ K). Of each sample, one half was doped 
with Gd, while the other half was left undoped.
Gd acts as a paramagnetic impurity and leads
to pair-breaking and a reduction of $T_{c}$. The transition
temperature as a function of Gd concentration was then studied as
a function of film thickness, which is correlated with $R_{\Box}$.
The remarkable result was that the pair-breaking parameter
$\alpha$ is only mildly dependent on disorder for films with
normal state sheet resistances $R_{\Box}$ ranging from 
$0.15\ {\rm k}\Omega$ to $2.2\ {\rm k}\Omega$. The implication seems to be
that the effects of disorder that lead to lower transition
temperatures do not manifest themselves in the spin-flip pair breaking 
rate. An attempt to understand this behavior by a phenomenological
modification of the Abrikosov-Gorkov result for $\alpha$,
using the renormalization of the density of states inherent in
Refs.\ \onlinecite{dbdos,db} and \onlinecite{trkdb}, failed.\cite{jj} 
This poses the important question whether the success of these theories in
describing the $T_c$ suppression and the disorder and temperature
dependence of the inelastic lifetime was fortuitous, and whether they
are lacking some important physical ingredient that manifests itself
in the pairbreaking rate.

It is the purpose of the present paper to analyze these questions.
What we will find is that one needs to take the strong-coupling
corrections\cite{am} to the Abrikosov-Gorkov expression into
account before one generalizes to the disordered case in order to obtain 
the correct structure of the theory. Once this is done, we find that
our previous theory\cite{db,tpddb} accounts very well for the
observed effect.

\section{formalism}
\label{sec:II}
Our starting point is our theory for the suppression of the
superconducting $T_c$,\cite{db}
and the enhancement of the inelastic scattering rate,\cite{tpddb}
by nonmagnetic disorder.\cite{TheoryChoice}
In this section we recall the most salient features of this theory.
First one uses an exact eigenstate formalism
to derive generalized Eliashberg equations for the normal Green
function $G(\epsilon,i\omega)$, and the anomalous Green function
$F(\epsilon,i\omega)$. Since the wavenumber is not a good quantum number
in the presence of static impurities, $G$ and $F$ are functions of
energy and frequency, rather than wavevector and frequency as in
Eliashberg theory. The Green functions are expressed, as usual, in terms
of an anomalous self-energy, $W(\epsilon,i\omega)$, and a normal 
one.\cite{ssw}
The latter is split into a piece $i\omega Z(\epsilon,i\omega)$ that is 
an odd function of frequency, and a piece $Y(\epsilon,i\omega)$ that
is even in $\omega$. In terms of these quantities, $G$ and $F$ read,
\begin{mathletters}
\label{eqs:1}
\begin{equation}
G(\epsilon,i\omega)=
{i\omega Z(\epsilon,i\omega)+\epsilon+Y(\epsilon,i\omega)\over{[i\omega 
Z(\epsilon,i\omega)]^{2}-[\epsilon+Y(\epsilon,i\omega)]^{2}}}\quad,
\label{eq:1a}
\end{equation}
and
\begin{equation}
F(\epsilon,i\omega)={-W(\epsilon,i\omega)\over{[i\omega 
Z(\epsilon,i\omega)]^2 - [\epsilon+Y(\epsilon,i\omega)]^{2}}}\quad.
\label{eq:1b} 
\end{equation}
\end{mathletters}%
In clean systems, the normal self-energy piece $Y$ is a constant that
just shifts the chemical potential and can be omitted. In the presence
of disorder, however, $Y$ has been found to be of crucial 
importance,\cite{db,blm,db2} and to reflect the physics of the Coulomb
anomaly in the density of states\cite{aa} in the context
of superconductivity.
The generalized Eliashberg equations then take the form of integral
equations in both energy and frequency for the three functions $Z$,
$W$, and $Y$. A solution of these equations has been obtained by
means of various approximations. In particular, the frequency dependence
of $Y$ was found to be weak and could be omitted, and its energy
dependence was approximated by the first term in a Taylor expansion
about a characteristic energy $\bar\omega$,
\begin{equation}
Y(\epsilon,i\omega) \simeq (\epsilon-\bar\omega) Y^{\prime}\quad,
\label{eq:2}
\end{equation}
with $Y^{\prime} = \partial Y/\partial\epsilon \vert_{\epsilon=\bar\omega}$, 
and $\bar\omega$ an average phonon
frequency.\cite{vdg,db2} With some further approximations, the
energy integrations could then be performed, and the theory be cast
in the same form as Eliashberg theory. A two-square well approximation
then leads to a $T_c$ formula that has the structure of a generalized
McMillan or Allen-Dynes formula, viz.\cite{db,db2}
\begin{equation}
T_c = {\omega_{\log}\over 1.2}\exp\left[{-1.04(1+\tilde\lambda+Y^{\prime})
        \over \tilde\lambda - \tilde\mu^{*}[1 + 0.62\tilde\lambda/(1+Y^{\prime})]}
         \right]\quad.
\label{eq:3}
\end{equation}
Here $Y^{\prime}$ is the normal self-energy piece mentioned above,
and $\tilde\lambda$ and $\tilde\mu^{*}$ are disorder dependent 
generalizations of the
electron-phonon coupling constant $\lambda$ and the Coulomb pseudotential
$\mu^{*}$, respectively, in Eliashberg theory. Explicit expressions for
all three of these quantities have been given in Ref.\ \onlinecite{db},
and will be evaluated for the case of thin films below.

In the presence of magnetic impurities, $T_c$ is reduced by pair
breaking.\cite{ag} Abrikosov-Gorkov theory has been modified to allow
for strong-coupling effects,\cite{am} with the only resulting change being
a factor of $1/Z$ in the pair breaking parameter. The result is
\begin{equation}
-\ln(T_{c}/T_{c0})=\psi(\alpha/2\pi T_{c}+1/2)-\psi(1/2)\quad,
\label{eq:4}
\end{equation}
with T$_{c0}$ the value of $T_c$ in the absence of the magnetic impurities,
$\psi$ the di-gamma function, and $\alpha$ the pair breaking
parameter $\alpha=(1/Z)(1/\tau_{s})$, with $1/\tau_{s}$ the spin-flip 
scattering rate. In the case of clean superconductors, $Z=1+\lambda$.
In the presence of nonmagnetic disorder, and with the same approximations
that lead to the $T_c$ formula given by Eq.\ (\ref{eq:3}), it is 
straightforward to repeat the calculation of Allen and Mitrovic\cite{am} 
within the framework of the theory of Ref.\ \onlinecite{db}. The
result is again Eq.\ (\ref{eq:4}), but with $\alpha$ replaced by a
disorder dependent $\tilde\alpha$ which in turn is related to
disorder dependent parameters,
\begin{equation}
\tilde\alpha={1/\tilde\tau_s\over{1+\tilde\lambda}}\quad.
\label{eq:5}
\end{equation}
Here $\tilde\lambda$ is the same quantity as in Eq.\ (\ref{eq:3}), and
$1/\tilde\tau_s$ is the disorder dependent spin flip-scattering rate.
Throughout this paper, we choose units such that $\hbar = k_B = 1$.
In the next section, we derive an explicit form for $\tilde\alpha$ in a
disordered thin superconducting film.

\section{disorder dependence of $\alpha$}
\label{sec:III}

\subsection{Electron-phonon coupling $\tilde\lambda$}
\label{subsec:III.A}

The electron-phonon coupling strength $\tilde\lambda$ is defined as
an integral over the Eliashberg function $\alpha^2 F$,
\begin{equation}
\tilde\lambda=2\int {d\nu\over{\nu}}\ \alpha^{2}F(\nu)\quad.
\label{eq:6}
\end{equation}
$\alpha^2 F$, and hence $\tilde\lambda$, are disorder dependent due
to effects first discussed by Pippard\cite{pippard} in the context
of ultrasonic attenuation, and by Schmid\cite{schmid} for the electron-phonon
inelastic lifetime. The main physical point is that disorder decreases
the coupling of the electrons to longitudinal phonons due to collision drag, 
but increases the coupling to transverse phonons due to the breakdown of
momentum conservation. For realistic parameter values the 
latter effect is stronger than the former, leading to an overall increase
of $\tilde\lambda$ with disorder.\cite{dba2F} For Debye phonons in $3-d$
systems, $\tilde\lambda$ has been calculated in Ref.\ \onlinecite{db}.
Repeating that calculation in $d=2$ is straightforward, and very similar
to the corresponding calculation of the electron-phonon inelastic 
lifetime.\cite{dbsds} The result is\cite{tpdthesis}
\begin{equation}
\tilde\lambda = 2\int_{0}^{\omega_{D}}{d\nu\over{\nu}} 
                          {\nu^{2}l\over{\pi m}}
        \sum_{b=L,T} {d_{b}\over{c_{b}^{3}}}\ f_{b}(\nu l/c_{b})\quad,
\label{eq:7}
\end{equation}
for a system with mean free path $l$. Here $c_{L,T}$ are the longitudinal and 
transverse speeds of sound, respectively, $\omega_{D}$ is the Debye frequency,
the dimensionless constant
$d_{b}=k_{F}^{3}/16\pi \rho_{i}c_{b}$ with
$\rho_{i}$ the ion density and $k_F$ the Fermi wave number, and
the functions $f_{T,L}$ are given by,\cite{dbsds}
\begin{mathletters}
\label{eqs:8}
\begin{equation}
f_{T}(x)={8\over{x^{4}}}(1+x^{2}/2-\sqrt{1+x^{2}})\quad,
\label{eq:8a}
\end{equation}
\begin{equation}
f_{L}(x)=2\left({1\over{\sqrt{1+x^{2}}-1}} -{2\over{x^{2}}}\right)\quad.
\label{eq:8b}
\end{equation}
\end{mathletters}
Substituting Eqs.\ (\ref{eqs:8}) into Eq.\ (\ref{eq:7}) we obtain the
disorder dependence of $\tilde\lambda$,
\begin{mathletters}
\label{eqs:9}
\begin{eqnarray}
\tilde\lambda&=&{\lambda\hat R_{\Box}E_{F} c_{L}\over{\pi\omega_{D}v_{F}}}
\biggl[F_{L}\left({\pi\omega_{D}v_{F}\over{\hat R_{\Box}E_{F}c_{L}}}\right)
\nonumber\\
&+&2{c_{L}^{2}\over{c_{T}^{2}}} F_{T}\left({\pi\omega_{D}v_{F}\over
{\hat R_{\Box}E_{F}c_{T}}}\right)\biggr] \quad,
\label{eq:9a}
\end{eqnarray}
where we have defined two functions,
\begin{eqnarray}
&F_{L}(x)&=\sqrt{1+x^{2}}-1-\ln[(\sqrt{1+x^{2}}+1)/2]\quad, \nonumber \\
&F_{T}(x)&= {1-\sqrt{1+x^{2}}\over{2(1+\sqrt{1+x^{2}})}}
+\ln[(\sqrt{1+x^{2}}+1)/2]\quad, 
\label{eq:9b}
\end{eqnarray}
\end{mathletters}%
with $\lambda = 4\omega_{D}d_{L}/\pi m c_{L}^{2}$ the electron-phonon
coupling in a clean $2-d$ system. The dimensionless resistance
$\hat R_{\Box} = R_{\Box} e^2/\hbar \approx R_{\Box}/4.1\ {\rm k}\Omega$ 
is a measure of the disorder in the material.

As in three dimensions, the size of the disorder renormalization of $\lambda$ 
depends on the ratio of the longitudinal to the transverse
speed of sound. This is a result of the abovementioned
competition between an increase in the coupling between electrons and
transverse phonons and a decrease of the
coupling to longitudinal phonons.
Since the transverse speed of sound is invariably smaller than the
longitudinal one, $\tilde\lambda$ increases with increasing disorder.
This effect tends to reduce the pair
breaking rate, Eq. (5). However, we also have to calculate the disorder
dependence of the spin-flip scattering rate in order to obtain the
disorder dependence of $\tilde\alpha$.

\subsection{Spin-flip scattering rate $1/\tau_{s}$}
\label{subsec:III.B}

The interaction between the electron spin and an impurity spin 
${\vec S}({\vec r})$ at site ${\vec r}$
is described by a Hamiltonian,
\begin{equation}
H_{S}=\sum_{{\bf k},{\bf k^{\prime}},\mu,\nu} J_{{\bf k},{\bf k^{\prime}}}\ 
{\vec S}({\bf k}-{\bf k^{\prime}})\cdot 
(c^{\dagger}_{{\bf k^{\prime}}\mu}\vec\sigma_{\mu\nu}c_{{\bf k}\nu})\quad.
\label{eq:10}
\end{equation}
Here $c^{\dagger}$ and $c$ are fermion operators, 
$\vec\sigma = (\sigma_x, \sigma_y, \sigma_z)$ denotes the Pauli matrices,
and $J_{{\bf k},{\bf k^{\prime}}}$ denotes 
the electron-magnetic impurity exchange
integral. We now calculate the electron self energy contribution,
$\Sigma$, due to this interaction in Born approximation. It is most 
convenient to do this in
an exact eigenstate representation, in analogy to the calculation of
the Coulomb self energy in Ref.\ \onlinecite{aalr}. The calculation
is straightforward, and we obtain
\begin{equation}
\Sigma(\epsilon,i\omega)=\int {d\epsilon^{\prime}\over{N_{F}}} 
G(\epsilon^{\prime},i\omega)\sum_{{\bf q}}V_{S}({\bf q})
F_{s}({\bf q},\epsilon-\epsilon^{\prime})\quad.
\label{eq:11}
\end{equation}
Here, $N_{F}$ is the free electron density of 
states per spin at the Fermi level.
We only retain the $s-$wave component of the interaction so that
$V_{S}({\bf q})=n_{P}S(S+1) J^2$, where $n_{P}$ is the concentration of 
paramagetic impurities, and J is a measure of the exchange interaction
strength.\cite{rkky} $G(\epsilon,i\omega)$
is the normal Green function in the superconductor and is given by
Eq.\ (\ref{eq:1a}). Finally, $F_s$ is the spin density analogue of the
density-density correlation function denoted by $F$ in 
Ref.\ \onlinecite{aalr}. If we work to lowest order in the electron-impurity
spin interaction, and neglect Coulomb and finite temperature effects in
$\Sigma$, then $F$ and $F_s$ are identical.

We obtain the spin-flip scattering rate $1/\tau_s$ from the self energy
$\Sigma$ by analytically continuing to real frequencies, 
$i\omega \rightarrow \omega + i0$, and going `on-shell', i.e. putting
$\epsilon = \omega$. For our purposes, we are interested in the
influence of spin-flip scattering on the superconducting $T_c$. The
physics that determines the latter is dominated by processes on a
frequency scale of $\bar\omega$, a typical phonon frequency. For the
same reason for which we take the parameter $Y^{\prime}$ in Eq.\ (\ref{eq:2})
at the frequency $\bar\omega$ we therefore define
\begin{equation}
1/\tau_s = -2{\rm Im}\Sigma(\bar\omega,i\omega\rightarrow\bar\omega + i0)
         \quad.
\label{eq:12}
\end{equation}

In a clean system, the spin-density correlation function $F_s(q,\omega)$
is frequency independent. In that case we recover from Eq.\ (\ref{eq:12})
the well known result\cite{ag,am}
\begin{equation}
{1\over\tau_s}=n_{P}S(S+1) J^2 4N_{F}\quad.
\label{eq:13}
\end{equation}
In a disordered system, $F_s$ is diffusive,\cite{aalr} and in the
Green function $G$ we must keep the self energy piece $Y^{\prime}$ as
discussed above. We thus obtain
\begin{mathletters}
\label{eqs:14}
\begin{equation}
{1\over\tilde\tau_s}={2 n_{P}S(S+1)
J^2\over{N_{F}[1+Y^{\prime}]}}\sum_{\bf q}
F_s\left({\bf q},\bar\omega{\tilde\lambda \over 1 + Y^{\prime}}
                               \right)\quad,
\label{eq:14a}
\end{equation}
where
\begin{equation}
F_s({\bf q},\omega) = g(q) {Dq^2\over \omega^2 + (Dq^2)^2}\quad,
\label{eq:14b}
\end{equation}
\end{mathletters}%
with $g(q)$ the Lindhard function, which for simplicity we replace
by $N_F \Theta(2k_F - q)$. Here $D$ denotes the normal phase spin
density diffusion coefficient, which in
the noninteracting electron approximation coincides with the mass
or charge diffusion coefficient, so $D=\pi/m\hat R_{\Box}$.
Performing the wavenumber integral in Eq.\ (\ref{eq:14a}) we
finally obtain
\begin{mathletters}
\label{eqs:15}
\begin{equation}
{1\over\tilde\tau_s} = {1\over\tau_s}
\left\{1+{1\over1+Y^{\prime}}
{\hat R_{\Box}\over{8\pi}}
\ln\left[1 + \left({8\pi\over\hat R_{\Box}}{\epsilon_F\over
\bar\omega^{*}}\right)^{2}\right]\right\}\quad,
\label{eq:15a}
\end{equation}
with
\begin{equation}
\bar\omega^{*} = \bar\omega{\tilde\lambda\over 1 + Y^{\prime}}
                                                         \quad,
\label{eq:15b}
\end{equation}
\end{mathletters}%
and $1/\tau_s$ given by Eq.\ (\ref{eq:13}). $1/\tilde\tau_s$ depends on
disorder both explicitly, and implicitly through $Y^{\prime}$. Our
final task is therefore to calculate the dependence of $Y^{\prime}$
on $\hat R_{\Box}$.

\subsection{Normal Self Energy Piece $Y^{\prime}$}
\label{subsec:III.C}

In order to calculate $Y^{\prime}$ we again have to repeat the
calculations of Ref.\ \onlinecite{db} in $d=2$. Both the
electron-electron and the
electron-phonon contributions to the self energy contribute to the
self energy piece $Y$. Performing a Taylor
series expansion in energy around $\epsilon=\bar\omega$ of Eq. (2.12) of
Ref.\ \onlinecite{db}, we obtain
\begin{equation}
Y^{\prime}(\bar\omega)=\delta U_{C}^{Y}(\bar\omega)+4\int {d\nu\over{\nu}}
\delta\alpha^{2}F^{H}(\bar\omega,\nu)\quad.
\label{eq:16}
\end{equation}
$\delta U_{C}^{Y}(\bar\omega)$, which describes the Coulomb contribution
to $Y^{\prime}$, is taken from Ref.\ \onlinecite{db},
\begin{eqnarray}
& &\delta U_{C}^{Y}(\bar\omega)={1\over{\pi N_{F}}}\sum_{\bf q}
g({\bf q}){Dq^{2}\over{(Dq^{2})^{2}+\bar\omega^{2}}}\times
\label{eq:17} \\
& &\left\{
V_{C}({\bf q})-{2\over{g({\bf q})^{2}}}\sum_{\bf k,p}
\sum_{\bf k^{\prime},p^{\prime}}g_{\bf k,k^{\prime}}({\bf q})
g_{\bf p,p^{\prime}}({\bf q})V_{C}({\bf k-p})\right\},\nonumber
\end{eqnarray}
with the statically screened Coulomb potential,
\begin{equation}
V_{C}({\bf q})={1\over{2N_{F}}}{\kappa\over{\kappa+q}}\quad, \ \ \ \ \ \ \hfil
\kappa=4\pi e^{2}N_{F}\quad.
\label{eq:18}
\end{equation}
Using the prescription for performing momentum sums of this type as 
described in Ref.\ \onlinecite{db}, the integrals can be done and yield
\begin{eqnarray}
\delta U_{C}^{Y}(\bar\omega)&=&{\mu \hat R_{\Box}\over{8\pi}}
\biggl[G\left({\hat R_{\Box}\bar\omega\over{8\pi E_{F}}}{4k_{F}^{2}\over{\kappa^{2}}},{\hat R_{\Box}\bar\omega\over{8\pi E_{F}}},{2k_{F}\over{\kappa}}\right)\nonumber \\&-&{2\over{\pi}} H\left({\hat R_{\Box}\bar\omega\over{8\pi E_{F}}},{2k_{F}\over{\kappa}}\right)\biggr]\quad,
\label{eq:19}
\end{eqnarray}
with the functions
\begin{eqnarray}
&G&(x,y,z)={z\over{1+x^{2}}}{1\over{\ln(1+z)}}
\nonumber\\
&\times&\biggl\{\ln\left[
{1+1/y^{2}\over{(1+z)^{4}}}\right]-\sqrt{{x\over{2}}}(1-x)\ln\left[
{1-\sqrt{2y}+y\over{1+\sqrt{2y}+y}}\right]\nonumber \\
&+&\sqrt{2x}(1+x)\tan^{-1}\left({\sqrt{2y}\over{y-1}}\right)
-2x\tan^{-1}(1/y)\biggr\}\quad; \nonumber \\
&H&(y,z)={z\over{\ln(1+z)}}\ln(1+1/y^{2}) {1\over{\sqrt{z^{2}-1}}}
\nonumber\\
&\times &\ln\left[{z+\sqrt{z^{2}-1}\over{z-\sqrt{z^{2}-1}}}\right]\quad.
\label{eq:20}
\end{eqnarray}
The Coulomb pseudopotential $\mu$ in $d=2$ is given by
\begin{equation}
\mu={\kappa\over{2\pi k_{F}}}\ln\left(1+{2k_{F}\over{\kappa}}\right)\quad.
\label{eq:21}
\end{equation}

As discussed in Ref.\ \onlinecite{db}, 
the phonon contribution to $Y^{\prime}$, which is given by the second
term on the right-hand side of Eq.\ (\ref{eq:16}), is related to 
a stress-stress correlation function and can be calculated in a similar
manner as $\delta U_{C}^{Y}$. One obtains
\begin{eqnarray}
&4&\int {d\nu\over{\nu}} \delta\alpha^{2}F^{H}(\bar\omega,\nu)=\\
& &\lambda\hat R_{\Box}{c_{L}E_{F}\over{2\pi^{2}v_{F}\omega_{D}}}
\sin^{-1}\left[{\omega_{D}v_{F}\over{4 E_{F}c_{L}}}\right]
\ln\left[1+\left({8\pi E_{F}\over{\hat R_{\Box}\bar\omega}}\right)^{2}\right]\quad.
\nonumber
\end{eqnarray}
Finally, both contributions can be collected to give
\begin{eqnarray}
& &Y^{\prime}(\bar\omega)=\hat R_{\Box}\biggl\{{\mu\over{8\pi}}
[G-{2\over{\pi}} H] \nonumber \\
&+&\lambda{c_{L}E_{F}\over{2\pi^{2}v_{F}\omega_{D}}}
\sin^{-1}\left[{\omega_{D}v_{F}\over{4E_{F}c_{L}}}\right]
\ln\left[1+\left({8\pi E_{F}\over{\hat R_{\Box}\bar\omega}}\right)^{2}\right]\biggr\}\ ,
\label{eq:23}
\end{eqnarray}
where $G$ and $H$ have the same arguments as in Eq.\ (\ref{eq:19}).
The enhancement of $Y^{\prime}$ with increasing disorder is
due to the opening of a correlation gap in the (normal state) 
density of states, 
and contributes to the decrease of T$_{c}$.\cite{db,db2}

\section{Final Result and Discussion}
\label{sec:IV}

We are now in position to collect our results and thus
obtain the disorder dependence of the magnetic
pair breaking rate. Substituting Eqs.\ (\ref{eqs:9}), (\ref{eqs:15}), and
(\ref{eq:23}) into Eq.\ (\ref{eq:5}) yields our final result,
\begin{equation}
{\tilde\alpha\over\alpha}=
{1+\lambda\over{1+\tilde\lambda}}
\left\{1+{1\over{1+Y^{\prime}
}}{\hat R_{\Box}\over{8\pi}}\ln\left[1+\left({8\pi E_{F}\over{\hat R_{\Box}\bar\omega^{*}}}\right)^{2}\right]\right\}.
\end{equation}
The disorder renormalizations of the pair breaking rate appear both
in the numerator and denominator and therefore the rate may either
increase or decrease with increasing 
disorder depending upon the material parameters
$\lambda,c_{L},c_{T},v_{F},E_{F},\omega_{D},$ and $\mu$. 

We now address the experiment on Pb$_{0.9}$Bi$_{0.1}$ by Chervenak
and Valles.\cite{jj}. 
To estimate the parameters entering into
Eq. (24), let us first consider the parameter values for bulk PbBi,
as far as available, given in Ref.\ \onlinecite{ad}. 
Thereby we have $\bar\omega=56 K$,
and $\omega_{D}=108K$. The Bohm-Staver relation gives  
\begin{equation}
{\omega_{D}\over{E_{F}}} {v_{F}\over{c_{L}}}=2 (2/Z)^{1/3}\quad.
\label{eq:25}
\end{equation}
For clean bulk Pb one has $E_{F}=1.1 \times 10^{5} K$, $c_{L}=$2050 m/s, $
c_{T}=710$ m/s, and $Z=4$. However, we do not expect the actual parameter
values to correspond to those for either bulk Pb or
bulk Pb$_{0.9}$Bi$_{0.1}$. First of all, the material in question is a thin
film, and moreover the substrate is expected to modify its properties,
in particular the acoustic ones. Evidence for this is provided by the fact
that the parameter values quoted above do not give 
the correct value of the clean limit 
$T_{c}$ as measured in Ref.\ \onlinecite{jj}.
It is therefore likely that the substrate on which the thin layer of
Pb is deposited strongly affects the phonon spectra of the film,
altering both $\lambda$ and the ratio $c_{L}/c_{T}$ compared to bulk Pb. 
Accordingly, we choose a value for the bare $\lambda=1.12$ and 
$\mu=0.1$ which (in the absence of disorder) reproduces the highest T$_{c}$ as 
measured in Ref.\ \onlinecite{jj}. Lastly, since $c_{L}/c_{T}$ is not known 
even for Pb$_{0.9}$Bi$_{0.1}$, we let $c_{L}/c_{L}$ be determined by a 
fit to the data.

These parameters provide the curves shown in Figure 1. The solid curves
are the results for the disorder dependence of the normalized pair
breaking rate as given by Eq. (5). The points represent the data taken from
Ref. \ \onlinecite{jj} for two different runs. The decrease
at small $\hat R_{\Box}$ is due to the fact that for small disorder,
$\tilde\lambda$ and $Y^{\prime}$ grow more rapidly 
than $1/\tilde\tau_{s}$. The normalized rate
goes through a shallow minimum at roughly $\hat R_{\Box}\sim 0.3$ at which
point the disorder renormalizations of $Y^{\prime},\tilde\lambda,$ and
$1/\tilde\tau_{s}$ are balanced and offset each other. With further
increasing disorder the enhancement of $1/\tilde\tau_s$ dominates,
and leads to a slowly increasing pair breaking rate.

We remark that the point at which the minimum occurs depends sensitively
on the ratio of the longitudinal and transverse speeds of sound. 
To obtain the solid lines in Fig. 1a, 
$c_{L}/c_{T}=1.9$ was used while 2.1 was used for Fig. 1b. 
These values lie between the
value for bulk Pb (2.88) and the substrate (similar to 
pyrex, 1.72) used in Ref. \ \onlinecite{jj} and thus does not seem 
unreasonable. Larger values 
of $c_{L}/c_{T}$ yield a more drastic reduction of the rate for small
disorder and the region of increasing $\alpha$
occurs at larger values of $R_{\Box}$. This sensitivity of the overall
shape of the curve to the material parameters may be reflected in the
slightly different results obtained for the two experimental runs in
Ref. \ \onlinecite{jj} as shown in Fig. 1. It would therefore be very 
interesting to repeat 
the experiments using substrates with different acoustic properties.

In summary, we have presented a theory for the paramagnetic pair breaking 
rate in disordered superconducting films and have shown that the
disorder dependence of the rate depends delicately on the disorder
renormalizations of $Y^{\prime}$, $\tilde\lambda,$ and 
$1/\tilde\tau_{s}$. As a result, the rate can
either increase or decrease with disorder, depending upon material parameters,
and in general it is not a monotonic function of disorder.
Our conclusion is that the disorder dependence of the rate as observed by
Chervenak and Valles in Pb$_{0.9}$Bi$_{0.1}$ films can be quantitatively
understood via an application of the
microscopic theories developed in Refs. \ \onlinecite{tpddb} and \ 
\onlinecite{db}.

\acknowledgements
We gratefully acknowledge helpful discussions with Jay Chervenak, 
Jim Valles, and Martin Wybourne. Part of this work was performed at
the TSRC in Telluride, CO, and we thank the Center for its hospitality.
This work was supported by the NSF under grant numbers DMR-92-06023,
DMR-92-09879, and DMR-95-10185.

\newpage
\begin{figure}[t]
\epsfxsize=3.5in
\epsfysize=3.in
\epsffile{jj.epsi}
\vskip 1.0cm
\end{figure}
\begin{figure}[t]
\epsfxsize=3.5in
\epsfysize=3.in
\epsffile{jj1.epsi}
\vskip 1.0cm
\caption{Plot of the pair breaking rate as a function of disorder
for two different Pb$_{0.9}$Bi$_{0.1}$ films. The symbols are data from
Ref.\ \protect\onlinecite{jj}, the solid line is obtained
from Eq.\ (\protect\ref{eq:5}). For Fig. 1a(b) we have chosen $c_{L}/c_{T}=1.9$ ($2.1$),
and $\alpha=1/\tau_s (1+\lambda)=90.5$K ($70$K). All
other parameters are the same for both figures and are
given in the text.}
\end{figure}

\begin{references}
\bibitem{TheoryChoice} We choose the theory of Refs.\ \onlinecite{db,tpddb}
 rather than the one of Ref.\ \onlinecite{trkdb}, even though the latter
 contains a somewhat more sophisticated treatment of the disorder and the
 resulting renormalization of the density of states. The reason is that
 strong-coupling effects, which were not included in Ref.\ \onlinecite{trkdb},
 are crucial for understanding the pair breaking rate.
\bibitem{rkky} Here we take $J$ to be a phenomenological
 coupling constant that is independent of disorder.
\end{references}
\end{document}